\documentclass[a4paper,12pt]{article}
\usepackage{graphicx, graphics}
\usepackage[T1,T2A]{fontenc}
\usepackage[utf8]{inputenc}
\usepackage{amssymb}
\usepackage{amsmath}
\usepackage{setspace}
\usepackage{xcolor}
\usepackage{wrapfig}
\graphicspath{{images/}}
\usepackage{multirow}
\usepackage{makecell}

\usepackage{bm}

\usepackage{siunitx}
\usepackage{booktabs}
\usepackage{array}
\newcolumntype{C}[1]{>{\centering\arraybackslash}p{#1}}

\begin{document} 

\thispagestyle{empty}

\begin{center}
{\large
{\bf 
Markov chains for the analysis of states of one-dimensional spin systems
} 
}
\vskip0.5\baselineskip{
\bf 

D.~N.~Yasinskaya$^{1}$,
Y.~D.~Panov$^{1}$
}
\vskip0.5\baselineskip{
$^{1}$Ural Federal University, 620002, Ekaterinburg, Russia
}
\end{center}

\begin{center}
\textbf{Annotation}
\end{center}

We analyze frustrated states of the one-dimensional dilute Ising chain with charged interacting impurities of two types with mapping of the system to some Markov chain. We perform classification and reveal two types of Markov chains: periodic with period 2 and aperiodic. Frustrated phases with various types of chains have different properties. In phases with periodic Markov chains, long-range order is observed in the sublattice while another sublattice remains disordered. This results in a conjunction of the non-zero residual entropy and the infinite correlation length. In frustrated phases with aperiodic chains, there is no long-range order, and the correlation length remains finite. It is shown that under the magnetic field the most significant change in the spin chain structure corresponds to the change of the Markov chain type.\\

The study was supported by a grant from the Russian Science Foundation, project  24-22-00196.\\


DOI: 10.61011/PSS.2024.07.58978.47HH

\newpage

\section{Introduction}

The one-dimensional spin models, despite their apparent
simplicity compared to the multi-dimensional models, have
a set of unique properties. The exact solutions for these
models underpin the understanding of a complex behavior
of real physical systems and play an important role in
studying such phenomena as phase transitions in statistical
physics~\cite{Baxter}. The absence or complexity of the long-range order formation underlies the unusual behavior of
the low-dimensional (pseudo)spin systems. The presence
of anisotropy and frustration in the system contribute to
a variety of phase diagrams and also to such unusual
phenomena as magnetic plateaus~\cite{Aydiner}, quasi-phases and
pseudo-transitions~\cite{Souza}, as well as the enhancement of the
magnetocaloric effect~\cite{Zhitomirsky}. The disorder also significantly
affects the phase, critical and magnetic properties of the
systems, and serves as a source of frustration. The frustrated
phases, provided that Rojas criterion is fulfilled~\cite{Rojas2020}, can
be the cause of such subtle pseudo-critical phenomenon
as pseudo-transitions expressed as a jump-like change of
the system disordered state and are accompanied by sharp
features of some thermodynamic functions.

Despite the availability of an exact solution, the analysis
of the phase states in one-dimensional systems within the
framework of standard formalism presents a non-trivial
challenge, especially for states at the boundaries between
different phases. An alternative approach in this case could
be the construction of a mapping of the one-dimensional
model onto a Markov chain, which has previously been
utilized to analyze the frustrated phase states of a dilute
Ising chain in a magnetic field~\cite{PanovPRE2022,PanovPSS2023}, as well as for Potts
model on a diamond chain~\cite{PanovPRE2023}. Such a mapping can be constructed for any model, which partition function
can be expressed via the transfer-matrix, that is true, for
instance, for various versions of Ising, Potts, Blume-Capel
and Blume-Emery-Griffiths models

One of the sources of frustration in 1D spin systems
is the introduction of impurities~\cite{PanovPSS2023}. In this work, we
consider the dilute Ising chain where the charged impurities
of two types are introduced. The 2D version of this model
was obtained and studied earlier as an atomic limit for a
pseudospin model of cuprates~\cite{Moskvin2013}. The ground state and
thermodynamic properties of the dilute Ising system are
influenced by both the frustration due to impurities and
the competition of the charge and magnetic orderings~\cite{Panov2019}.
Through numerical simulations on a square lattice, it has
been shown that this leads to the presence of non-universal
critical behavior~\cite{YasinskayaIEEE2021}, first-order phase transitions~\cite{Yasinskaya2021}, and
reentrant phase transitions~\cite{Yasinskaya2020}.

The article is organized as follows. Section 2 includes the constructed and studied phase diagrams of the
1D Ising model with charged impurities in the variables
``exchange interaction parameter -- chemical potential''.
Also, concentration dependences are found for the residual
entropy of the frustrated phases. Section 3 outlines
methodology of mapping the one-dimensional model onto
a Markov chain; expressions for the transition matrix,
equilibrium state and correlation functions are provided.
The Markov chain properties and correlation properties for
the frustrated ferromagnetic and antiferromagnetic phases
were analyzed. In Section 4, the types of Markov chains
are classified according to their symmetry and correlation
properties; the influence of the magnetic field on the
Markov chains was reviewed. A summary is provided in
Section 5.

\section{Ground state phase diagrams and
	residual entropy of the
	frustrated phases}

The ground state phase diagrams, as well as temperature
phase diagrams of 2D Ising model with two types of nonmagnetic impurities on a square lattice were calculated
earlier in works~\cite{Panov2019, Yasinskaya2020} by mean field method and computational modeling in zero magnetic field at specified charge
density of non-magnetic impurities n, as one of the system
parameters.

The Hamiltonian of a 1D dilute Ising model is expressed
as follows

\begin{equation}
	\mathcal{H} = \sum_{i=1}^N \left\{ \Delta  S_{z,i}^2 + V S_{z,i} S_{z,i+1} + J P_{0,i} s_{z,i} s_{z,i+1} P_{0,i+1} - h P_{0,i} s_{z,i} - \mu S_{z,i} \right\},
\label{H}
\end{equation}
where $\Delta$ -- one-site density-density correlations in the form
of a single-ion anisotropy for the pseudospin $S = 1$; $V$ --
inter-site density-density interaction; $J$ -- Ising exchange
interaction of spins $s = 1/2$; $h$ -- external magnetic field;
$P_{0,i} = 1 - S_{z,i}^2$ -- projection operator for magnetic states.
The summation is carried out for $N$ sites of the chain.
Using the chemical potential $\mu$, a constraint is imposed
on the system in the form of conservation of total charge,
which can be expressed as fixing the charge density of
non-magnetic impurities: $n = \left\langle\sum_{i} S_{z,i} \right\rangle / N$. A detailed
discussion of the density-density interaction within the
pseudospin formalism was given in work~\cite{Yasinskaya2021}. Thus,
each site of the chain can be in one of the charge states
(spinless states of the pseudospin $S_z = \pm 1$ for positively
and negatively charged impurities, respectively), or in one of
the spin states (states of spin $s_z = \pm 1/2$, which correspond
to projection of the pseudospin $S_z = 0$).

Expressions for a grand thermodynamic potential of the
system per one site for different phases of the ground state
can be written as follows:

\begin{equation}
	\omega_{\text{I}}^{\pm} = \Delta + V \pm \mu, \, \, \omega_{\text{CO}} = \Delta - V, \,\, \omega_{\text{FM}}^{\pm} = J \pm h, \,\, \omega_{\text{AFM}} = -J, \,\, \omega_{\text{PM}}^{\pm\pm} = \frac{\Delta \pm h \pm \mu}{2}.
	\label{gpot}
\end{equation}
Impurity (I), ferromagnetic (FM), anti-ferromagnetic
(AFM), charge-ordered (CO) and paramagnetic (PM)
phases correspond to the following configurations
of the nearest neighbors for $h>0$: I$^{\pm}$ $\rightarrow$ $(\pm 1,\pm 1)$, FM $\rightarrow$ $\left(\frac{1}{2}, \frac{1}{2}\right)$, AFM $\rightarrow$ $\left(\frac{1}{2}, -\frac{1}{2}\right)$, CO $\rightarrow$ $(1,-1)$, PM$^{\pm}$ $\rightarrow$ $\left(\pm 1, \frac{1}{2}\right)$. 
 These ``pure'' phases are characterized
by the following values of the impurities charge density: $n_{\text{I}^{\pm}} = \pm 1$, $n_{\text{FM}} = n_{\text{AFM}} = n_{\text{CO}} = 0$, $n_{\text{PM}^{\pm}} = \pm \frac{1}{2}$.

By minimizing the grand potential of the system, new
phase diagrams may be built in the variables $(J,\mu)$.
2D areas in this case will correspond to the edge values in
terms of $n$ for the diagrams built in the representation with
the given $n$. On the contrary, the boundaries between the areas on phase diagrams in variables $(J,\mu)$ will correspond
to ``mixed'' phases with an intermediate value of $n$, that may
have non-zero residual entropy.

\begin{figure}[t]
	\centering
	\includegraphics[width=0.75\linewidth]{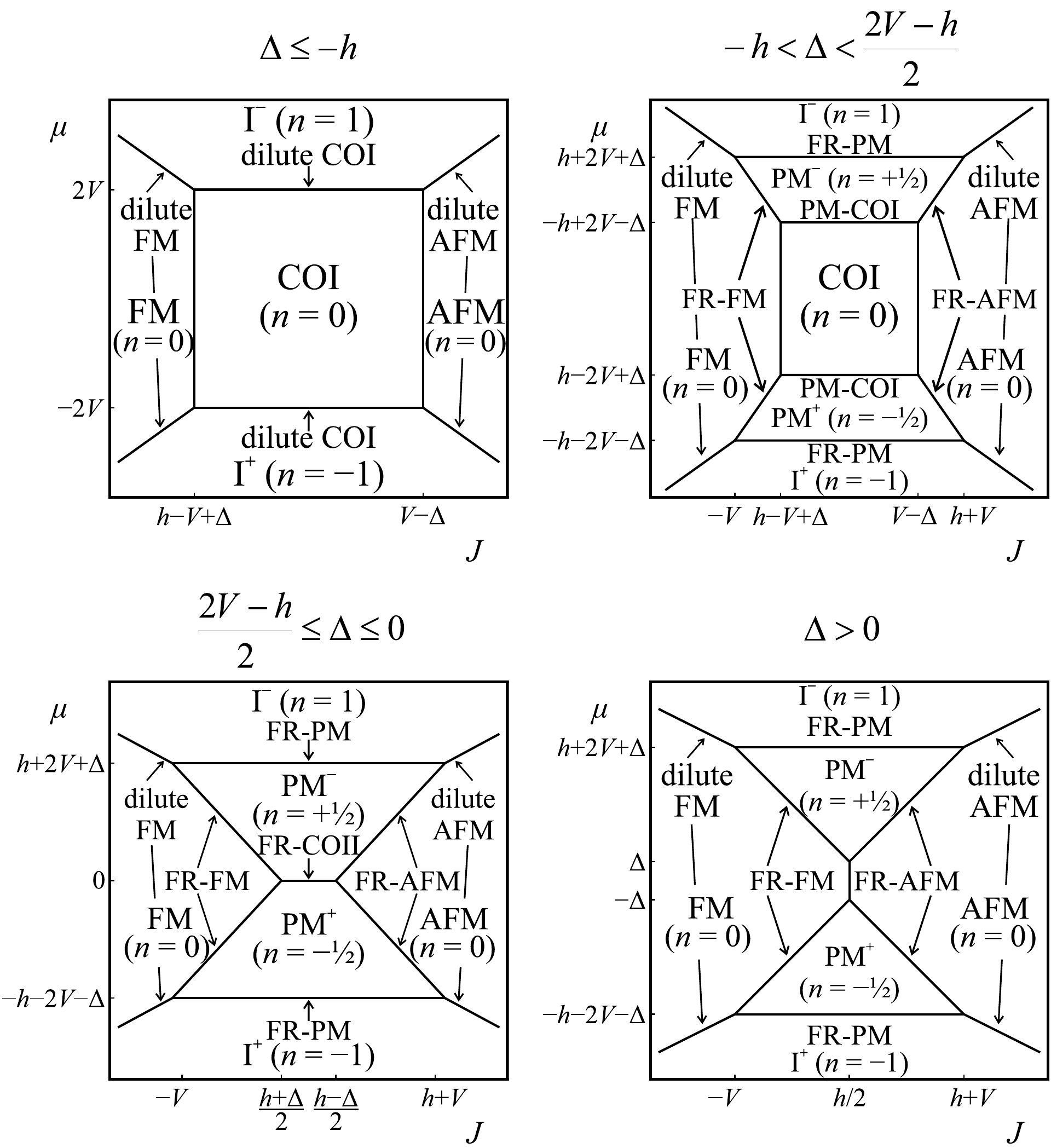}
	\caption{Ground state diagrams for $h \geq 2V$ (strong magnetic field) in variables $(J,\mu)$}
	\label{muPDhg2V}
\end{figure}

In a strong magnetic field ($h \geq 2V$), the four types of
phase diagrams are possible, as shown in Figure~\ref{muPDhg2V}. Thus,
for large negative $\Delta$, the ``pure'' phases are (A)FM, COI
(true for $n = 0$) and $I^{\pm}$ ($n = \pm 1$). Thus, the intersection
of (A)FM-phase with the impurity phase I will give a
dilute (anti)ferromagnetically ordered phase (dilute (A)FM)
with phase separation into magnetic domains and charged
droplets — macroscopic areas with a total volume $|n|$,
containing only sites filled with impurities. The number
of permutations of charged droplets in a chain that do not
change the ground state energy has a power-law asymptotic
behavior, and as one approaches the thermodynamic limit
$N \rightarrow \infty$, the residual entropy of these phases will tend
to zero. The intersection of COI with I is the phase
of the dilute charge ordering (dilute COI). The charged
impurities of one type are randomly distributed against the
background of a checkerboard charge ordering. There is an
exponentially large number of permutations of ``excessive''
impurities without the energy change, which leads to nonzero residual entropy. Thus, dilute COI-phase is frustrated.
The charge density of the impurities for both dilute phases
can take any value: $0 < |n| < 1$.

At $\Delta = -h$, a ``pure'' paramagnetic phase PM$^{\pm}$
($n = \pm \frac{1}{2}$)
appears, which at the interface with COI causes the charge
paramagnetic phase PM-COI that has non-zero residual
entropy and exist for $0 < \vert n \vert < \frac{1}{2}$. PM-COI is a dilute
checkerboard charge order with paramagnetic centers as
single spins, which are oriented along the field in the ground
state. The boundaries of (A)FM with PM provide frustrated
(anti) ferromagnetic phase FR-(A)FM with $0 < \vert n \vert < \frac{1}{2}$.
This is a dilute (A)FM-phase with (anti) ferromagnetically
ordered clusters (or single spins, aligned with magnetic
field) separated by single non-magnetic impurities with the
charge density of $n$. Here, in contrast to the dilute (A)FM phase, the non-magnetic impurities are not collected into
a charged droplet, but are distributed randomly throughout
the entire system, resulting in a non-zero residual entropy.
At $\frac{2V-h}{2} \leq \Delta \leq 0$ an additional boundary between the
two paramagnetic phases PM$^+$ and PM$^-$ appears, which
corresponds to the frustrated charge phase FR-COII with
sublattice-alternating spins aligned with the field and nonmagnetic impurities of both charges. This phase occurs at
$0 < \vert n \vert < \frac{1}{2}$.

To obtain the expressions for residual entropy of frustrated phases within the ``standard'' transfer-matrix approach, it is necessary to find the maximum eigenvalue of
the transfer-matrix, determine the parametrical dependence
of the entropy from the charge density using the chemical
potential, and find the limit at the zero temperature.
This is quite a sophisticated task; however, based on the
Markov property of the dilute Ising chain~\cite{PanovJMMM}, one can
analytically define the concentration dependencies of the
residual entropy of all frustrated phases of the ground state
by an alternative method~\cite{PanovPRE2022}.

\begin{figure}[h!]
	\centering
	\includegraphics[width=\linewidth]{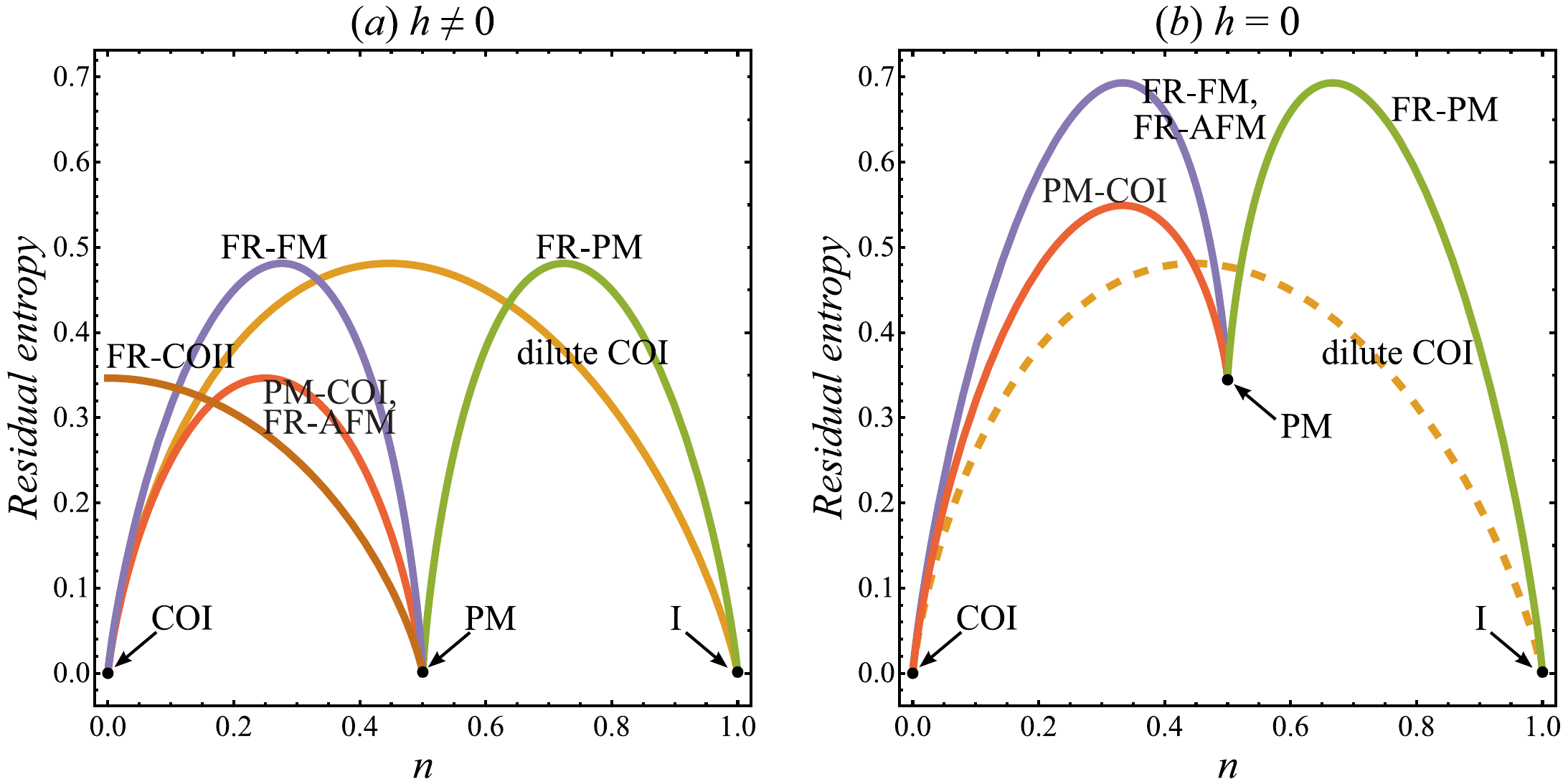}
	\caption{Concentration dependencies of the residual entropy of various ground state phases at (a) $h \neq 0$; (b) $h=0$}
	\label{s(n)}
\end{figure}

The dependences of the residual entropy of the ground
state phases on the charge density of impurities n are
given in Figure~\ref{s(n)} in zero (b) and non-zero (a) magnetic
fields. The dilute (A)FM phase has zero residual entropy
at all values of the charge density n and is not shown
in the Figure. The dilute COI phase has a non-zero
entropy for $n \neq 0$, which doesn’t depend on magnetic
field; it reaches a maximum value of $\frac{1}{2} \ln \frac{\sqrt{5} + 1}{\sqrt{5} - 1} \approx 0.481$ at $|n| = \frac{1}{\sqrt{5}} \approx 0.447$. 
The PM-COI and FR-AFM phases
in zero field have identical residual entropies, as they are
symmetric with respect to the replacement of spin states
with pseudospin states. The maximum entropy of these
phases is equal to $\frac{\ln 2}{2} \approx 0.347$, and is reached at $|n| = \frac{1}{4}$, while at $h = 0$ maximum values reach $\frac{\ln 3}{2}$ and $\ln 2$ at $|n| = \frac{1}{3}$ for PM-COI and FR-AFM, respectively. 
The entropies of
the FR-FM and FR-PM phases are symmetrical relative to
the edge point  $|n| = \frac{1}{2}$, and reach their maximum values
of $\frac{1}{2} \ln \frac{\sqrt{5} + 1}{\sqrt{5} - 1} \approx 0.481$ at $|n| = \frac{5 \mp \sqrt{5}}{10} \approx \frac{1}{2} \mp 0.224$ for $h \neq 0$ and $\ln 2$ at $|n| = \frac{1}{2} \mp \frac{1}{6}$ at $h=0$. 
The FR-COII phase is
of particular interest, it arises only in the strong magnetic
field, $h \geq 2V$. Residual entropy of this phase is non-zero and
reaches its maximum value of $\frac{\ln 2}{2} \approx 0.347$ at $n=0$, which
may be caused by re-arrangement of charge states without
the change of energy. This is the only phase that remains
frustrated at $n=0$.

It can be noted that the maximum values of the entropy
of the frustrated phases FR-COII, PM-COI, FR-AFM are
lower than those for dilute COI, FR-FM, FR-PM. Moreover,
in zero magnetic field, the entropies of FR-FM, FR-AFM,
PM-COI, FR-PM phases increase even more. The reason
for this lies in their structure, a detailed analysis of which
can also be carried out using a mapping of a 1D system
onto a Markov chain~\cite{PanovPSS2023,PanovPRE2023}. Such a mapping may be
simulated for any model, the partition function of which
allows presentation via the transfer-matrix, that is true, for
instance, for various versions of Ising, Potts, Blume-Capel
and Blume-Emery-Griffiths models

\section{Mapping a 1D spin model onto a
	Markov chain}

1D system transfer-matrix with Hamiltonian~\eqref{H} in the
local basis of states 
$$\Phi = \left\{ \left|S_z, s_z\right\rangle \right\} = \left\{\left|+1,0\right\rangle, \left|-1,0\right\rangle, \left|0,+\frac{1}{2}\right\rangle, \left|0,-\frac{1}{2}\right\rangle \right\} \equiv \left\{+1, -1, +\frac{1}{2}, -\frac{1}{2} \right\}$$ has a structure
\begin{equation}
	\hat{T} = \begin{pmatrix}
		e^{-\beta \omega^{-}_{\text{I}} } & e^{-\beta \omega_{\text{CO}}} & e^{-\beta \omega^{--}_{\text{PM}} } & e^{-\beta \omega^{+-}_{\text{PM}} } \\
		e^{-\beta \omega_{\text{CO}}}  & e^{-\beta \omega^{+}_{\text{I}} } & e^{-\beta \omega^{-+}_{\text{PM}} } & e^{-\beta \omega^{++}_{\text{PM}} } \\
		e^{-\beta \omega^{--}_{\text{PM}} } & e^{-\beta \omega^{-+}_{\text{PM}} } & e^{-\beta \omega^{-}_{\text{FM}} } & e^{-\beta \omega_{\text{AFM}} } \\
		e^{-\beta \omega^{+-}_{\text{PM}} } & e^{-\beta \omega^{++}_{\text{PM}} } & e^{-\beta \omega_{\text{AFM}} } & e^{-\beta \omega^+_{\text{FM}} } \\
	\end{pmatrix},
	\label{TM}
\end{equation}
where the notations~\eqref{gpot} are used for grand thermodynamic
potentials of the ground state phases.

As elements of the transition matrix of the Markov
chain, one can take the conditional probabilities $P(b \vert a)$
of realizing state $b$ at the $(i + 1)$-site given that the
$i$-site is in state $a$. 
The conditional probabilities can
be defined by Bayes formula $P(ab) = P(a) P(b \vert a)$, where
$P(a) = \langle P_{a,i} \rangle$ -- probability of implementation of a state at the $i$-site, $P(ab) = \langle P_{a,i} P_{b,i+1} \rangle$ -- probability of joint
implementation of $a$ and $b$ states on $i$- and $(i + 1)$-sites,
respectively, $P_{a,i}$ -- projector on $a$ state for siteа $i$.

By using the transfer matrix~\eqref{TM} built on the states $a$, one
can find the correlators~\cite{PanovPRE2023}
\begin{equation}
	\label{corr1}
	\langle P_{a,i} \rangle = \langle a \vert \lambda_1 \rangle \langle \lambda_1 \vert a \rangle
\end{equation}
\begin{equation}
	\langle P_{a,i} P_{b,i+l} \rangle = \langle a \vert \lambda_1 \rangle \frac{T_{ab}^l}{\lambda_1^l} \langle \lambda_1 \vert b \rangle,
\end{equation}
where $\lambda_1$ -- the largest eigenvalue of the transfer-matrix, $\langle a \vert \lambda_1 \rangle = \upsilon_a$ -- the element of the eigenvector of the
transfer-matrix for state $a$ corresponding to the largest
eigenvalue $\lambda_1$.

Thus, the conditional probabilities are equal to the
correlators ratio
\begin{equation}
	\label{pitab}
	P(b \vert a) = \frac{\langle P_{a,i} P_{b,i+l} \rangle}{\langle P_{a,i} \rangle} = \frac{T_{ab} \upsilon_b}{\lambda_1 \upsilon_a} = \pi_{ab}.
\end{equation}

Equilibrium (stationary) state of the system can be
expressed as a limiting distribution $\bm{p}$ of the Markov chain,
which remains unchanged as a result of the transition matrix
action. Respectively, for $p_a$ components of the limiting
distribution the following is true
\begin{equation}
	\sum_a p_a \pi_{ab} = p_b, \quad \sum_a p_a = 1, \quad
	p_a = P(a) = \langle P_{a,i} \rangle.
\end{equation}
Given~\eqref{corr1}, for symmetric transfer-matrices the limiting
distribution is associated with the normalized eigenvector
corresponding to the largest eigenvalue
\begin{equation}
	p_a = \upsilon_a^2.
	\label{statprob}
\end{equation}
Pair distribution functions can also be expressed using the
transition matrix, if we use the conclusion of Kolmogorov-
Chapman theorem:
\begin{equation}
	\langle P_{a,i} P_{b,i+l} \rangle = \sum\limits_{s_1, \dots , s_{l-1} } P(a) P(a \vert s_1) P( s_1 \vert s_2) \dots P(s_{l-1} \vert b) = p_a \pi_{ab}^l = \pi_{ba}^l p_b.
	\label{pairdistr}
\end{equation}
The correlation function for the states $a$ and $b$, thus, can
be expressed as follows:
\begin{equation}
	K_{ab}(l) = \langle P_{a,i} P_{b,i+l} \rangle - \langle P_{a,i} \rangle \langle P_{b,i} \rangle = p_a \pi_{ab}^l - p_a p_b = p_b \pi_{ba}^l - p_a p_b.
\end{equation}

Taking into account  $\sigma_{z,i} = P_{0,i} s_{z,i}/s = P_{\frac{1}{2},i} - P_{-\frac{1}{2},i}$,  one
can calculate the spin correlation function:
\begin{equation}
	C(l) = \langle \sigma_{z,i} \sigma_{z,i+l} \rangle - \langle \sigma_{z,i} \rangle^2 . 
\end{equation}

To construct the transition matrix and the corresponding
Markov chain for a specific frustrated ground state phases,
one can leave only the leading order elements in the transfer
matrix, and neglect the remaining elements due to their
exponentially small contributions at low temperatures. We
will do it for the frustrated magnetic phases FR-AFM and
FR-FM.

Let’s consider the system in the external magnetic field: $h > 0$. For convenience, we will examine the cases of
positive impurities charge: $n > 0$. 
Since FR-AFM phase
arises on the intersection of AFM and PM phases on
$(J,\mu)$-diagram (see Figure~\ref{muPDhg2V})), in the transfer-matrix we’ll
keep only the members corresponding to the necessary
configurations of the neighboring states:
$$
\hat{T} = \begin{pmatrix}
	0 & e^{-\beta \omega^{--}_{\text{PM}} } & 0\\
	e^{-\beta \omega^{--}_{\text{PM}} } & 0 & e^{-\beta \omega_{\text{AFM}} } \\
	0 & e^{-\beta \omega_{\text{AFM}} } & 0\\
\end{pmatrix} = \begin{pmatrix}
0  & e & 0\\
e  & 0 & d \\
0  & d & 0\\
\end{pmatrix}.
$$
$-1$  state is absent, the system states space will be reduced to $\Phi = \left\{ +1, +\frac{1}{2}, -\frac{1}{2} \right\}$. The largest eigenvalue of the transfermatrix is equal to $\lambda_1 = \sqrt{d^2 + e^2}$ with the eigenvector $\bm{v} = \left(	\frac{e}{\sqrt{2} \lambda_1},
0,
\frac{1}{\sqrt{2}},
\frac{d}{\sqrt{2} \lambda_1}\right)^T$.

According to expressions~\eqref{pitab},~\eqref{statprob}  let’s define the form of
the transition matrix and limiting distribution for this phase
\begin{equation}
	\pi = \begin{pmatrix}
		0 &  1 & 0 \\
		\frac{e^2}{\lambda_{1}^2} &  0 & \frac{d^2}{\lambda_{1}^2} & \\
		0 & 1 & 0 \\		
	\end{pmatrix},
	\qquad \bm{p} = \frac{1}{2 \lambda^2_{1}}\begin{pmatrix}
		e^2\\
		\lambda_{1}^2\\
		d^2\\
	\end{pmatrix}.
\end{equation}
The condition of constant charge density of impurities can
be written a
\begin{equation}
	n = P(1) - P(-1) = p_1 - p_{-1}.
\end{equation}

Then, the elements of the transition matrix and limiting
distribution can be expressed through $n$:
\begin{equation}
	\pi = \begin{pmatrix}
		0 & 1 & 0\\
		2n&0&1-2n\\
		0 &1&0\\
	\end{pmatrix}, \qquad \bm{p} = \begin{pmatrix}
		n\\
		\frac{1}{2}\\
		\frac{1}{2}-n\\
	\end{pmatrix}.
\end{equation}

\begin{figure}
	\centering
	\begin{minipage}{0.49\linewidth}
		\centering
		а)\\
		\includegraphics[height=5cm]{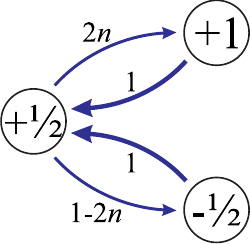}
	\end{minipage}
	\begin{minipage}{0.49\linewidth}
		\centering
		б)\\
		\includegraphics[height=5cm]{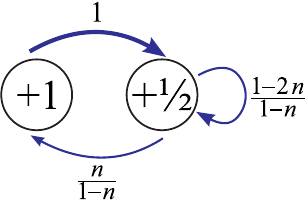}
	\end{minipage}
	\caption{Transition graphs between the states of Markov chain
		for the transition matrix of phases a) FR-AFM; b) FR-FM.}
	\label{fr-(a)fm-scheme}
\end{figure}

For the FR-AFM phase, the equilibrium state is the state
when the half of the chain is filled with spin $+\frac{1}{2}$, while 
the remaining part of the system is a mixture of positively
charged impurities with density of $n$ (with pseudospin +1) and spins $-\frac{1}{2}$ with density of $\frac{1}{2} - n$.  Based on the type
of transition matrix it is convenient to built the graph of
possible transitions. The vertices of the graph designate
possible states of the system, the links from one vertex to
another show possible transitions between the states. It is
presented in Figure~\ref{fr-(a)fm-scheme},а for the FR-AFM phase. Thicker
lines correspond to larger conditional probabilities of the
transition. Transitions to th $+\frac{1}{2}$ 
state from others occur
with probability 1, this state fully fills one sublattice, while
the second sublattice is chaotically filled with $+1$ and $- \frac{1}{2}$ 
states in accordance with a fixed value of $n$.

The impurity and spin correlation functions are equal,
correspondingly
\begin{equation}
	K_{+1,+1}(l) = (-1)^l n^2,\quad
	C(l) = (-1)^l (1-n)^2 .
\end{equation}
Both correlation functions are characterized by the infinite
correlation length: $\xi = \infty$.

Thus, the chain is divided into two sublattices. One of
them is fully ordered — filled with $+\frac{1}{2}$
spins, which gives an
infinite correlation length. In the second sublattice, the spins
$-\frac{1}{2}$
are replaced by $+1$ impurities with increasing $n$ and are
arranged chaotically. The state of this sublattice is frustrated
and is characterized by the correlation functions equal to
zero. Because of this, the FR-AFM phase combines ordering
on one sublattice and chaotic character on the other, which
gives infinite correlation length and non-zero entropy. It is
clearly seen when analyzing the two-step transition matrix:
\begin{equation}
\pi^2 = \begin{pmatrix}
	2 n & 0 & 1-2 n \\
	0 & 1 & 0 \\
	2 n & 0 & 1-2 n \\
\end{pmatrix}
\end{equation}

The state space of the two-step Markov chain splits into two
independent sub-spaces: $\Phi = \left\{ +\frac{1}{2} \right\} \bigcup \left\{ +1, - \frac{1}{2} \right\}$, which
describe two sublattices of the spin chain.

As a result, FR-AFM phase may be represented as a
set of AFM-clusters separated by single impurities. In this
case, AFM clusters always contain an odd number of spins
and have states $+\frac{1}{2}$
at the edges, aligned with the external
magnetic field $h > 0$. 

Let’s consider an equivalent frustrated ferromagnetic
phase FR-FM in the field $h > 0$. It corresponds to the boundary between phases FM and PM on $(J,\mu)$-diagram
(see Figure~\ref{muPDhg2V}). The transfer-matrix, transition matrix and
equilibrium system state in this phase will be expressed as
follows:
\begin{equation}
	\hat{T} = \begin{pmatrix}
	0 & e^{-\beta \omega^{--}_{\text{PM}} }\\
	e^{-\beta \omega^{--}_{\text{PM}} } & e^{-\beta \omega_{\text{FM}}^+ }\\
\end{pmatrix}, \quad
\pi = \begin{pmatrix}
	0 & 1 \\
	\frac{n}{1-n}&\frac{1-2n}{1-n}\\
\end{pmatrix}, \quad \bm{p} = \begin{pmatrix}
	n\\
	1-n
\end{pmatrix}.
\end{equation}
Now the states space is reduced to $\Phi = \left\{ +1, +\frac{1}{2} \right\}$. The
system equilibrium state is itself the ferromagnetic clusters
separated by single non-magnetic impurities. The transition
graph is shown in Figure~\ref{fr-(a)fm-scheme},b.

The impurity and spin correlation functions are equal to:
\begin{equation}
	K_{+1+1}(l) = K_{+\frac{1}{2}+\frac{1}{2}}(l) = C(l) = (-1)^l n(1-n) e^{-l/\xi},
\end{equation}
where the correlation length $\xi$ is finite and depends on the
charge density:
\begin{equation}
	\xi = \left[ \ln \left(\frac{1-n}{n}\right) \right]^{-1}.
\end{equation}
This means, that at $h \neq 0$ there are no critical fluctuations,
and the state remains frustrated and disordered even at
$T=0$. The correlation length is equal to zero at $n = 0$,
when the FR-FM phase transits into the ordered FM-phase
with entropy equal to zero.

In the absence of external magnetic field the properties
of FR-AFM and FR-FM phases will slightly change. For the
FR-AFM phase the mutual relations between the impurities
state $+1$ and spin state $-\frac{1}{2}$ will be added into Markov
chain. For the FR-FM phase the spin state $-\frac{1}{2}$, having
the same relation with impurities, as $+\frac{1}{2}$
state, will appear.
Now the spin states $\pm\frac{1}{2}$
are included in Markov chain
symmetrically. The correlation functions in the zero field
decrease exponentially for both phases,
\begin{equation}
	K_{+1+1}^{\text{FR-AFM}}(l) = K_{+1+1}^{\text{FR-FM}} = (-1)^l (1-n)n e^{-l/\xi_{c}},
\end{equation}
\begin{equation}
	C^{\text{FR-AFM}}(l) = (-1)^l C^{\text{FR-FM}}(l) = (-1)^l (1-n) e^{-l/\xi_s},
\end{equation}
where charge and spin correlation lengths are equal, correspondingly
\begin{equation}
	\xi_{c} = \left[ \ln \left(\frac{1-n}{n}\right) \right]^{-1}, \quad
	\xi_s = \left[ \ln \left(\frac{1-n}{1-2n}\right) \right]^{-1}.
\end{equation}
As a result, the application of an external magnetic field
may induce a subtle rearrangement of states, leading to the
emergence of long-range order characterized by an infinite
correlation length within one of the sublattices. In contrast,
in the ferromagnetic phase FR-FM the response to the
magnetic field is primarily characterized by the flipping of
spin clusters in the direction of the applied field.

Thus, the analysis of the ground state phases using
Markov chains facilitates the identification of the features
of the phases structure and reveals the hidden sublattice
ordering. This method also allows for the analytical determination of correlation functions and correlation lengths, as
well as the calculation of residual entropy.

\section{Types of Markov chains
	of the 1D dilute Ising model}

Now we classify the types of Markov chains and determine their form for the existing phases. The results for the
frustrated ground state phases are presented in the Table~\ref{marktab},
where for each phase we provide the transition matrix $\pi$,
the form of the equilibrium state $\bm{p}$, the transition graph
between states in the space $\Phi$, as well as the form of the
correlation functions and correlation lengths.

\begin{table}
	\caption{Таблица отображений фрустрированных фаз на марковские цепи ($n>0$)}
	\label{marktab}
	\small
	\begin{tabular}{|>{\centering\arraybackslash}m{1.4cm}|>{\centering\arraybackslash}m{1.25cm}|>{\centering\arraybackslash}m{3.35cm}|>{\centering\arraybackslash}m{1.35cm}|>{\centering\arraybackslash}m{1.7cm}|>{\centering\arraybackslash}m{4.2cm}|}
		\hline
		Type of Markov chain & Phase & Transition matrix $\pi$ & Equilibrium state $\bm{p}$ & Transition graph & Correlation functions $K_{+1+1}(l)$, $C(l)$ and correlation lengths \\
		\hline
		\hline
		\multirow{3}{*}{\rotatebox[origin=c]{90}{\parbox[c]{4cm}{\centering Periodic with a period of 2}}} & FR-AFM ($h > 0$) & $\begin{pmatrix}
			0 & 1 & 0\\
			2n&0&1-2n\\
			0 &1&0\\
		\end{pmatrix}$ & $ \begin{pmatrix}
			n\\
			\frac{1}{2}\\
			\frac{1}{2}-n\\
		\end{pmatrix}$ & 
	 \includegraphics[width=1.7cm]{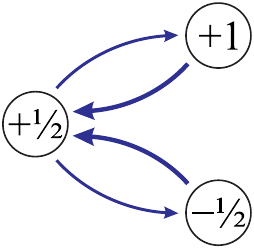} & \vspace{-0.5cm} \makecell{$K_{+1+1}(l) = 
		(-1)^l n^2$,\\ $C(l) = (-1)^l (1-n)^2$,\\ $\xi = \infty$}  \\
		\cline{2-6}
		& PM-COI ($h>0$) & $\begin{pmatrix}
			0 & 1-2n & 2n \\
			1 & 0 & 0 \\
			1 & 0 & 0\\
		\end{pmatrix}$ & $ \begin{pmatrix}
			\frac{1}{2}\\
			\frac{1}{2} - n\\
			n
		\end{pmatrix}$ & 
		\includegraphics[width=1.7cm]{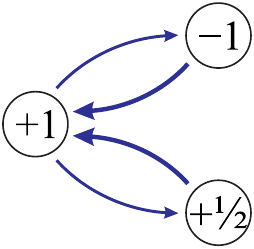}  & 
		 \vspace{-0.5cm} \makecell{$K_{+1+1}(l) = \frac{(-1)^l}{4}$,\\ $C(l) = (-1)^l n^2 $, \\ $\xi = \infty$} \\
		\cline{2-6}
		& FR-COII & $\begin{pmatrix}
			0 & 0 & 1 \\
			0 & 0 & 1\\
			\frac{1}{2} + n & \frac{1}{2} -n & 0\\
		\end{pmatrix}$ & $\begin{pmatrix}
			\frac{1+2n}{4} \\
			\frac{1-2n}{4} \\
			\frac{1}{2}
		\end{pmatrix}$ & 
	 \includegraphics[width=1.7cm]{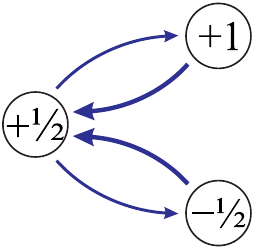} & \vspace{-0.5cm}\makecell{$K_{+1+1}(l) =  \frac{(-1)^l (1+2n)^2}{16}$,\\ $C(l) = \frac{(-1)^l}{4}$, \\ $\xi = \infty$} \\
		\hline
		\hline
	\multirow{3}{*}{\rotatebox[origin=c]{90}{\parbox[c]{4cm}{\centering
				Aperiodic}}} &	FR-FM ($h > 0$) & $\begin{pmatrix}
			0 & 1 \\
			\frac{n}{1-n}& \frac{1-2n}{1-n}\\
		\end{pmatrix}$ & \multirow[b]{2}{*}[-0.3cm]{$\begin{pmatrix}
			n\\
			1-n
		\end{pmatrix}$} & 
	 \includegraphics[width=1.7cm]{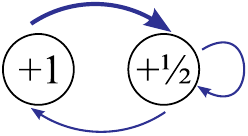} & $K_{+1+1}(l) = C(l) = (-1)^l n(1-n) e^{-l/\xi}$, $\xi = \left[ \ln \left(\frac{1-n}{n}\right) \right]^{-1}$  \\
		\cline{2-3}\cline{5-6}
		& FR-PM ($h>0$) & $\begin{pmatrix}
			\frac{2n-1}{n} & \frac{1-n}{n}\\
			1 & 0 \\
		\end{pmatrix}$ &  & 
	 \includegraphics[width=1.7cm]{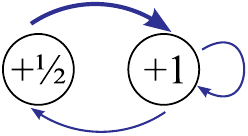} & $K_{+1+1}(l) = C(l) = (-1)^l n(1-n) e^{-l/\xi}$, $\xi = \left[ \ln \left(\frac{n}{1-n}\right) \right]^{-1}$  \\
		\cline{2-6}
		& dilute COI & $\begin{pmatrix}
			\frac{2n}{1+n} & \frac{1-n}{1+n} \\
			1 & 0 \\
		\end{pmatrix}$ & $\begin{pmatrix}
			\frac{1+n}{2}\\
			\frac{1-n}{2}
		\end{pmatrix}$ & 
	 \includegraphics[width=1.7cm]{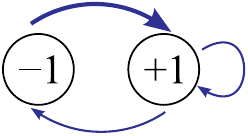} & $K_{+1+1}(l) = \frac{(-1)^l}{4} \left(1 - n^2\right) e^{-l/\xi}$, $\xi = \left[ \ln \left(\frac{1+n}{1-n}\right) \right]^{-1}$  \\
		\hline
		\hline
			\rotatebox[origin=c]{90}{\parbox[c]{2.4cm}{\centering Periodic with a period of 2}} & PM-COI ($h=0$) & $ \begin{pmatrix}
			0 & 1-2n & n & n \\
			1 & 0 & 0 & 0 \\
			1 & 0 & 0 & 0 \\
			1 & 0 & 0 & 0 \\
		\end{pmatrix}$ & $\begin{pmatrix}
			\frac{1}{2}\\
			\frac{1}{2}-n \\
			\frac{n}{2} \\
			\frac{n}{2}
		\end{pmatrix}$ & 
		\vspace{0.1cm} \includegraphics[width=1.7cm]{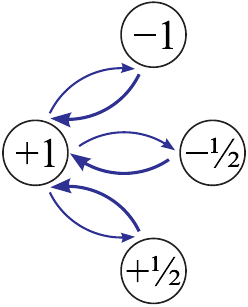}   &  \makecell{$
			K_{+1,+1} = \frac{(-1)^l}{4} $, \\ $\xi_c = \infty$,\\ $ C(l) = 0$} \\
		\hline
		\multirow{3}{*}{\rotatebox[origin=c]{90}{\parbox[c]{4cm}{\centering Aperiodic}}} & FR-AFM ($h=0$) & $ \begin{pmatrix}
			0 & \frac{1}{2} & \frac{1}{2} \\
			\frac{n}{1-n} & 0 & \frac{1-2n}{1-n} \\
			\frac{n}{1-n} & \frac{1-2n}{1-n} & 0
		\end{pmatrix}$ & \multirow[b]{3}{*}[-1cm]{$\begin{pmatrix}
			n \\
			\frac{1-n}{2} \\
			\frac{1-n}{2}
		\end{pmatrix}$} & 
	 \includegraphics[width=1.7cm]{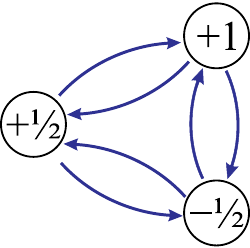} &  \multirow{2}{*}{\makecell{$
				K_{+1,+1} = $ \\$ (-1)^l (1-n) n e^{-\l/\xi_c}$, \\ $\xi_c = \left[ \ln \left(\frac{1-n}{n}\right) \right]^{-1}$, \\ 
				$ C(l) = (\mp 1)^l (1-n) e^{-l/\xi_s}$,\\ 
				$ \xi_s = \left[ \ln \left(\frac{1-n}{1-2n}\right) \right]^{-1}
				$}} \\
		\cline{2-3}\cline{5-5}
		& FR-FM ($h = 0$) & $\begin{pmatrix}
			0 & \frac{1}{2} & \frac{1}{2} \\
			\frac{n}{1-n} & \frac{1-2n}{1-n} & 0 \\
			\frac{n}{1-n} & 0 & \frac{1-2n}{1-n}
		\end{pmatrix}$ &  & 
	 \includegraphics[width=1.7cm]{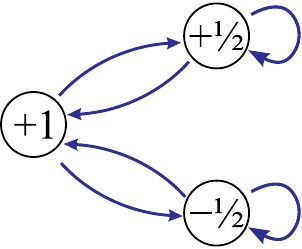} &  \\
	\cline{2-3}\cline{5-6}
	 & FR-PM ($h=0$) & $ \begin{pmatrix}
		\frac{2n-1}{n} & \frac{1-n}{2n} & \frac{1-n}{2n} \\
		1 & 0 & 0 \\
		1 & 0 & 0
	\end{pmatrix}$ &  & 
\includegraphics[width=1.7cm]{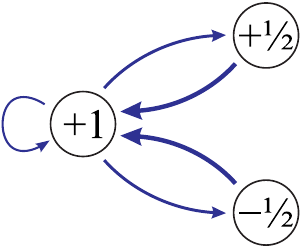} & $
	K_{+1,+1} = (-1)^l n e^{-\l/\xi}$, $\xi_c = \left[ \ln \left(\frac{n}{1-n}\right) \right]^{-1}$, $ C(l) = 0$  \\
	\hline
	\end{tabular}
\end{table}

In the presence of a magnetic field, we can distinguish
two types of Markov chains; they are given in the first two
parts of the Table~\ref{marktab}. The FR-AFM, FR-COII and PM-COI
phases have Markov chains with a period of 2 and an infinite
correlation length due to the ordered sublattice. As a result,
the residual entropy for these phases is lower than that of
the second class of frustrated phases. The second class
phases, FR-FM, dilute COI, and FR-PM, are characterized
by a Markov chain consisting of two states and a finite
correlation length that depends on $n$.

The rearrangements of Markov chains at $h = 0$ also
occur in other phases sensitive to a magnetic field. The
characteristics of Markov chains for such phases in the
absence of a magnetic field are given in the last part of the
Table~\ref{marktab}. The Markov chain for the paramagnetic charge phase
PM-COI in the absence of a magnetic field retains a period
of 2; however, now, in addition to the state $+\frac{1}{2}$, the opposite
spin state $-\frac{1}{2}$
emerges. While the long-range ordering of
impurities in the sublattice remains stable at any magnetic
field, the spin states become uncorrelated in the absence of a
magnetic field. Markov chain of the frustrated paramagnetic
phase FR-PM in zero field contains an additional state $-\frac{1}{2}$,
which is symmetric to $+\frac{1}{2}$. The charge correlation length
remains unchanged; however, the spin correlation function
becomes zero, which increases the residual entropy of this
phase.

\begin{figure}[h!]
	\centering
	\includegraphics[width=0.6\linewidth]{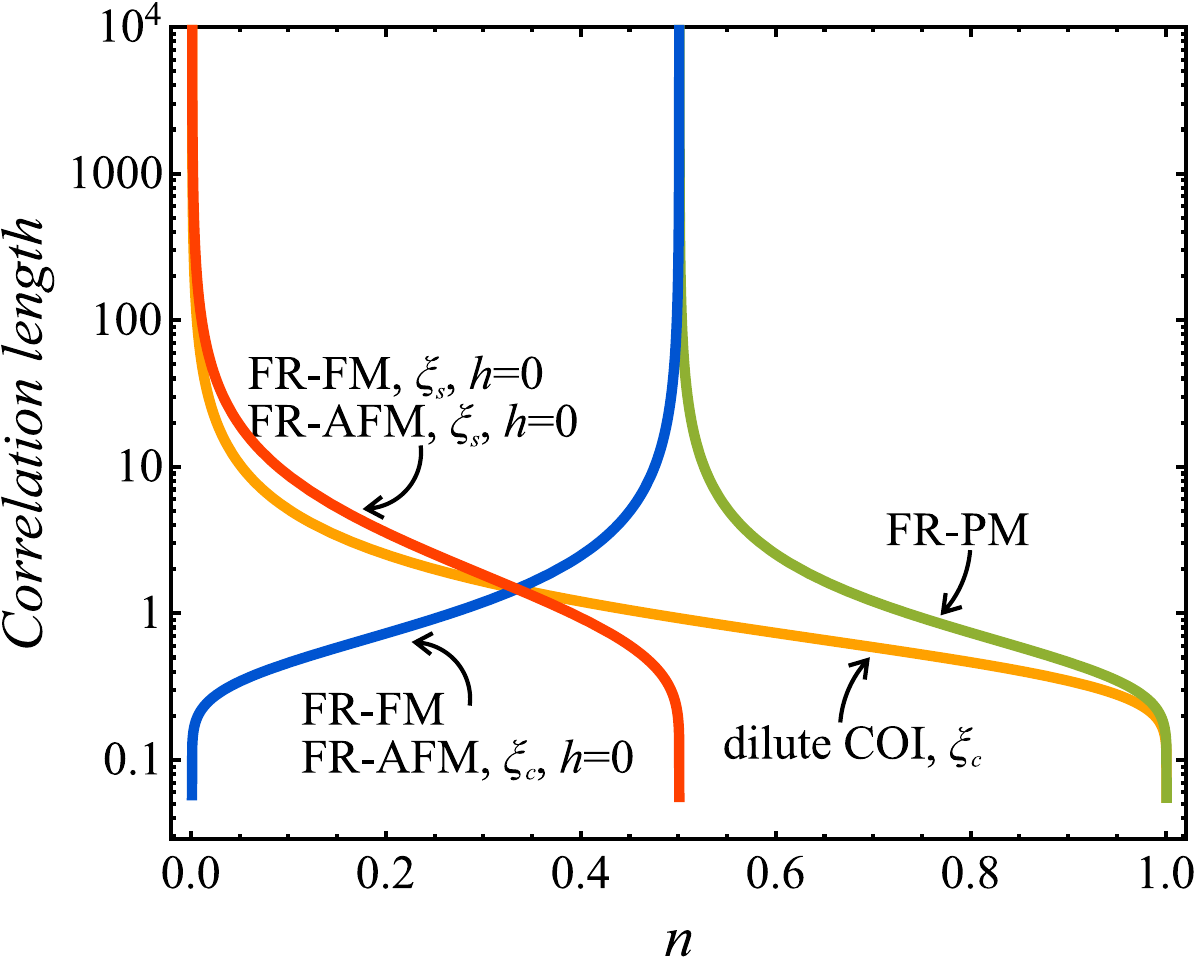}
	\caption{Correlation length of ground state phases as a function
		of charge density of non-magnetic impurities $n$ in a logarithmic
		scale.}
	\label{corr(n)}
\end{figure}

The Figure~\ref{corr(n)} illustrates the concentration dependencies
of the correlation lengths of various frustrated phases in a
logarithmic scale, both in the presence and absence of a
magnetic field $h$. The correlation lengths of phases with
periodic Markov chains are not depicted, since they are
infinite for any $n$.

The impurity correlation length $\xi_c$ for the phases
FR-(A)FM at $h = 0$ is the same as that for the FR-FM
phase at $h \neq 0$. It becomes zero at $n = 0$, when frustrated
phases transition into ``pure'' (A)FM phases. In that case,
the spin correlation length $\xi_s$ goes to infinity. This indicates
a magnetic phase transition at $T = 0$, when the (A)FM-state
becomes fully ordered. Another boundary point is $n = \frac{1}{2}$,
corresponding to ``pure'' paramagnetic ordering PM. In this
point the impurity correlation length is divergent, while the
spin correlation length tends to zero. In the FR-PM phase,
compared to the FR-FM, the roles of impurity and spin
states are interchanged, resulting in symmetrical properties
of these phases relative to the $n = \frac{1}{2}$
point; absence of
the spin correlation in the FR-PM phase in zero magnetic
field is an exclusion. In the dilute COI phase the long-range checkerboard charge ordering at $T = 0$ occurs only
in ``pure'' COI limit, at $n = 0$, when the charge correlation
length becomes divergent.

\section{Conclusion}

Using a method based on Markov chains analysis, the
properties of frustrated phases of the Ising chain with two
types of charged impurities have been studied.

The considered model exhibits a wide variety of ground
state phases, most of which have non-zero residual entropy,
and tend to be frustrated in this context. In zero magnetic
field each frustrated phase has its own type of Markov chain.
In an external field, the system has only 2 types of Markov
chains, each of which is typical for three different frustrated
phases of the ground state. The properties of these 2
Markov chains differ significantly. Thus, the frustrated
antiferromagnetic FR-AFM phase, the paramagnetic charge
phase PM-COI and a mixture of impurity paramagnetic
phases FR-COII have periodic Markov chains with a period
of 2 and three states. This indicates the presence of
ordering on one of the spin chain sublattices, while the
second sublattice remains completely disordered. Due to
this hidden ordering, the correlation length of the system is
infinite, while the residual entropy is relatively small. The
frustrated ferromagnetic FR-FM, checkerboard charge dilute
COI and paramagnetic FR-PM phases can be effectively described by aperiodic Markov chains with two states. This
is in line with a spin chain composed of clusters of one state
type, which are separated by single sites of the second state
type. In this case, no long-range order is established in the
system, resulting in a finite correlation length that depends
on the charge density. Consequently, the residual entropy is
higher than that of the first-type phases.

The performed analysis shows that when a magnetic field
is included, the most significant change in the structure of
the spin chain corresponds to a change in the type of the
Markov chain: for the frustrated antiferromagnetic phase
FR-AFM, the aperiodic Markov chain becomes periodic,
which indicates the emergence of a long-range order in the
system

\section{Funding}

This study was supported financially by grant No. 24-22-
00196 from the Russian Science Foundation.

\makeatletter

\end{document}